# TALYS calculation and a short review of the experimental status of proton capture studies on p-nuclei: A guide to future investigation*

Indrani Ray† Argha Deb
Physics Department, Jadavpur University, Kolkata - 700 032, India

**Abstract:** TALYS calculations were performed to obtain the theoretical proton capture cross-sections on the *p*-nuclei. A short review on the status of related experimental studies was also conducted. Some basic properties such as *Q*-values, Coulomb barrier, Gamow peak, Gamow Window, and decay properties of the parent and daughter nuclei were studied. Various experimental parameters, e.g., beam energy, beam current, targets, and detectors, used in experimental investigations reported in the literature, were tabulated. The results of the TALYS calculations in the Gamow region were compared with the corresponding experimental values wherever available. This study is expected to facilitate the planning of future experiments.

## I. INTRODUCTION

Hydrogen and helium are the two most abundant elements in the Universe and were formed in the primordial or Big Bang nucleosynthesis. The stellar nucleosynthesis starts with a sequential reaction called the pp1-chain [1−3], which involves fusion of H to form He. Once the H fuel becomes exhausted, He starts burning to form the next most abundant elements, namely carbon and oxygen. The next important process is the CNO cycle [1−3], whose end product is the same as that of the pp1 chain. After He burning stops, depending on the stellar temperature and mass, the different successive phases of C, O, Ne, and Si burning take place until iron is produced. The charged particle reactions are the primary processes in nucleosynthesis till mass $A \approx 60$ is reached, above which neutron capture through s (slow) and r (rapid) processes [1−3] becomes the dominant process. It has been observed that among the heavier elements ($A > 60$), there is a small fraction of proton-rich nuclei which are 10−1000 times less abundant than s- or r-nuclei and cannot be synthesized through these neutron capture processes because they are on the neutron deficient side of the $\beta$-stability line. This group, consisting of approximately 35 nuclei between Se ($Z$=34) and Hg ($Z$=80), is synthesized by a different astrophysical process traditionally called the *p*-process or $\gamma$-process, and these nuclei are referred to as the *p*-nuclei [1−6]. Various studies indicate that the lighter $p$ nuclei might be dominantly produced in the r$p$ and $\nu p$ processes and that the heavier $p$ nuclei are synthesized through a sequence of photodisintegrations on the seeds of *r*- and *s*-process nuclei [7]. Recent studies suggest that some of the $p$ nuclei can also be produced in charged particle reactions [7].In recent years, both theoretical and experimental investigations have been conducted to understand the $p$ process nucleosynthesis [5−7]. To solve the problem of the origin of the *p*-nuclei, it is essential to make use of both astrophysical and nuclear models as well as the available experimental information. Precise measurements enable constraining our models, and well tested nuclear models can provide an estimate of how reliable the measurements are. The experimental data are still inadequate because the production of these nuclei seems to require several processes. The present study was undertaken with the intention of providing a useful guide to plan future experiments for measuring proton capture cross-sections on the *p*-nuclei using gamma-ray spectroscopy as a tool. For proper planning of experiments and careful analysis of their outcome, it is essential to have a clear knowledge of quantities such as *Q*-value, Coulomb barrier, Gamow energy, and decay properties of the system under study. It is also crucial to make a careful review of the experimental techniques and methods of analysis adopted by various researchers to extract useful and reliable data. This study is an attempt to summarize such information in a concise format to be readily available for scientists aiming to experimentally investigate proton capture cross-sections on the *p*-nuclei. The theoretical calculations were performed using TALYS code. These results, when compared with existing experimental data, help identify the systems which require reinvestigation or those for which no or scarce experimental data exist.

## II. IMPORTANT MATHEMATICAL FORMULAE AND METHODOLOGY

Nuclear reaction rates are essential ingredients for investigation of energy generation and nucleosynthesis processes in stars. Astrophysical reaction rates describing the change in the abundances due to nuclear reactions taking place in an astrophysical environment are functions of the densities of the interacting nuclei, their relative velocities, and reaction cross-sections [6, 8]. The role of nuclear physics in understanding the process of nucleosynthesis and stellar evolution involves measurements or computations of fundamental quantities such as reaction cross-sections, nuclear masses, and $\beta$-decay rates. As mentioned in the previous section, the majority of $p$ nuclei are synthesized through a sequence of photodisintegration reactions. In principle, photodisintegration cross-sections can be experimentally determined by photon induced reaction studies [9, 10]. However, in an experiment performed in laboratory, the target nucleus is always in the ground state, whereas in a stellar environment, thermally populated excited states significantly contribute to the reaction rate. It has been shown that the influence of thermal population is much less pronounced in capture reactions [11]. In addition, there is a lack of availability of appropriate $\gamma$-sources. Therefore, it is advantageous to measure the capture cross-sections; the photodisintegration rates can be derived by applying the principle of detailed balance [3]. The various important capture reactions to understand the $p$-process nucleosynthesis are $(n, \gamma)$, $(p, \gamma)$, $(\alpha, \gamma)$, *etc*. In the present study, we focused on the proton capture $(p, \gamma)$ reaction. ($\Delta E_0$) The charged particle induced nuclear reactions take place within a narrow energy window called Gamow window around the effective burning energy known as Gamow peak $E_0$, which lies well below the Coulomb energy $E_c$. The Gamow peak arises from folding the Coulomb penetration probability with the Maxwell-Boltzmann velocity distribution and is expressed as

$$E_0 = \left(\frac{bkT}{2}\right)^{\frac{3}{2}} \tag{1}$$

where, $b = 31.28 Z_1 Z_2 A^{1/2}$ keV$^{1/2}$ is the barrier penetrability. The corresponding expression for Gamow window is

$$\Delta E_0 = \frac{4}{\sqrt{3}} (E_0 k_B T)^{1/2} \tag{2}$$

Ideally, nuclear astrophysics experiments should be performed within the energy region given by $E_0 \pm \Delta E_0$. The cross-section for charge particle reactions for $E < E_c$ ($E$ is the center of mass energy for the reacting system) is

$$\sigma = S(E)\frac{1}{E}\exp\left(-2\pi\eta(E)\right) \tag{3}$$

which includes relevant information on nuclear physics. The factor $\eta$ is known as the Sommerfield parameter and is expressed as $\eta(E) = \frac{Z_1 Z_2 e^2}{\hbar v}$, where $Z_1$ and $Z_2$ are the atomic numbers of the interacting nuclei, $v$ is their relative velocity, and $\hbar = \frac{h}{2\pi}$. The reaction rate for a pair of interacting nuclei is given by

$$\langle \sigma v \rangle = \left(\frac{8}{\pi\mu}\right)^{1/2} \frac{1}{k_B T}^{3/2} \int_0^\infty S(E) exp\left(-\frac{E}{k_B T} - 2\pi\eta(E)\right) dE \tag{4}$$

In the above expressions, $k_B$ is the Boltzman constant, $\mu = \frac{m_1 m_2}{m_1 + m_2}$ is the reduced mass of the interacting particles, and $T$ is the stellar temperature.

The importance of any particular nuclear reaction depends on the temperature and density of the specific stellar environment. Hence, each nuclear reaction must be treated as a unique process. The energy region of the chosen nuclear reaction process depends on the relevant stellar environment. The reaction cross-section is measured by the number of reactions taking place and can be determined using various methods. Although the most common method is based on the detection of the light outgoing particle, which is an in-beam method, it is seldom adopted [12] for measuring particle capture cross-sections of astrophysical importance because reaction cross-sections are extremely small and difficult to separate from the beam induced background. The other possibility is the determination of the number of heavy residual nuclei, which requires a special experimental technique based on combining a recoil mass separator and inverse kinematics [13]. The third method is often conducted when the heavy residual nucleus is radioactive. This is the activation method [14], in which the cross-section is determined by detecting the decay or activity from the residual nucleus.

Measuring cross-sections in the astrophysically relevant region (Gamow window) region (Gamow window) is often impossible for various reasons. In many cases, low energy beams are not available. Even if it is possible to achieve a beam at the required energy, the cross-sections can be so small that it is generally impossible to separate the targeted data from the natural and cosmic-ray induced background. Hence, experiments are often performed at a higher energy and extrapolation provides the cross-section at the relevant astrophysical energies. In a simple case, the cross-section drops drastically with energy without exhibiting any significant structures. In this scenario, comparatively few points will be enough to constrain theoretical cross-section calculations. Sometimes, the extrapolation process becomes complicated owing to the presence of resonances, and the Breit-Wigner [15] formula is used to extrapolate the cross-sections to the region where no experimental data exist. The present study considers only non-resonant systems.

## III. PRESENT STUDY

This study is an attempt to provide guidance for future experimental studies involving *p*-nuclei. Neutron, proton, and alpha induced reactions play key roles in many astrophysical processes including the *p*-process nucleosynthesis. Very few charged–particle reactions have been experimentally investigated above the iron region. In general, investigations of gamma rays emitted from the radioactive nuclide have a very wide application scope for cross-section measurements. This is owing to the fact that gamma rays emitted from most radionuclides have a wide range of energies (40−1000 keV) with relatively high penetrating power, thereby ensuring minimal losses by absorption in a sample during measurement. This property, along with recent developments in high-resolution and high-efficiency gamma-ray detectors, makes gamma-ray spectroscopy a powerful technique for nuclear cross-section measurements. The present study includes a short review of experimental studies on the cross-sections of proton captures on *p*-nuclei through detection of gamma rays. TALYS calculations were performed to obtain theoretical estimates of these cross-section values (discussed in Sec. III.A). The current status of the corresponding experimental studies was reviewed following those reported in the KADoNis database [16]. These experimental results were compared with theoretical predictions from the TALYS calculations perfomed.

### A. Discussion of essential quantities for experimental planning

The values of the Coulomb barrier ($E_c$), $Q$-values, Gamow peak ($E_0$), and Gamow window ($\Delta E_0$) are represented against the corresponding mass numbers in Fig. 1 (a) − (c)). The points representing $E_c$ have been joined by lines to facilitate visual analysis. The Coulomb barrier was obtained from the fusion evaporation code PACE4 [17]. The $Q$-values were obtained from the NNDC website [18]. The values of the Gamow peaks and Gamow windows were calculated using Eqs. (1) and (2), respectively. The value of $T$ was set to $3\times10^{-9}$ K (T9 =3) to calculate the Gamow peak and Gamow window.

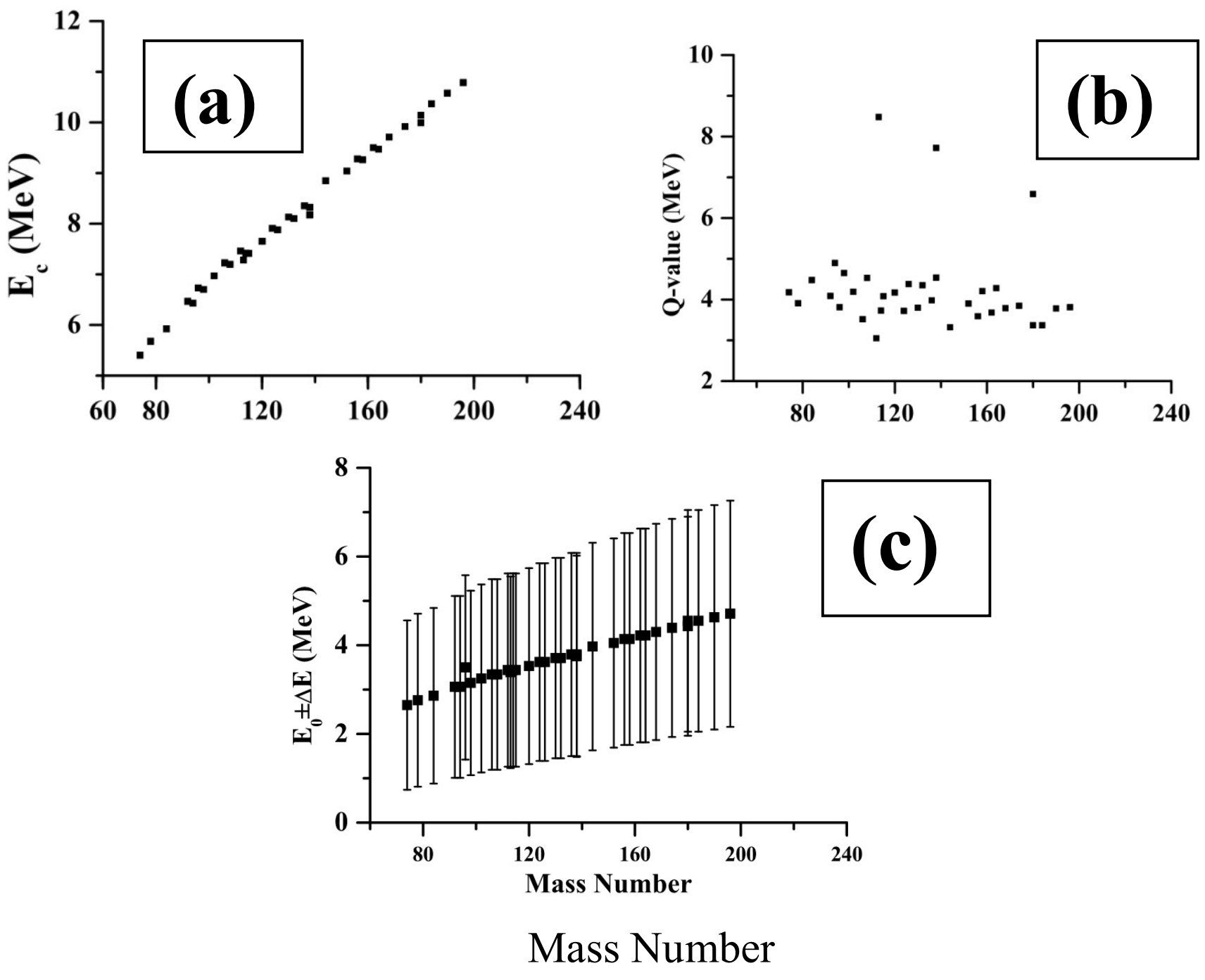

**Fig. 1:** Plots of Coulomb barrier ($E_c$), $Q$-values, Gamow peak ($E_0$), and Gamow window ($\Delta E_0$) against mass number.

The knowledge of lifetime values of the daughter nucleus is essential to choose a proper method for detection of gamma rays. For example, if the nucleus produced after proton capture has a stable ground state, then the in-beam technique is the only possible investigation method. However, if the ground state has a moderate half-life, activation analysis is the preferred investigation method. The values of the abundances or half-lives ($T_{1/2}$) of the target nuclei [19], half-lives ($T_{1/2}$) of the ground state of the daughter nucleus following proton capture, and a few significant gamma transitions of the nucleus formed following the subsequent decay ($\beta$, EC, IT) are listed in Table 1. These values were obtained from the NNDC website [20]. Some experimental parameters adopted by various authors are listed in Table 2. The references corresponding to the experimental studies reported in the literature are included in Table 1.

**TABLE 1:** Reactions, decay sequences**,** abundances of the target nuclei**,** lifetimes of the ground state of the daughter nuclei, significant gamma-decay lines of the final nucleus after decay (ec, β, IT) of the daughter nucleus, and references of the experimental studies.

| **Reaction and Decay** | **Abundance (NNDC) Target** | ***$T_{1/2}$* (NNDC) Target Daughter** | **γ - transitions feeding ground state (>2% of $I_{max}$) (keV)** | **Reference from KADoNis [10]** |
|---|---|---|---|---|
| p+$^{74}$Se→$^{75}$Br (EC)→$^{75}$Se | 0.89 | 96.7min | 112,287,428 | 21 |
| p+$^{78}$Kr→$^{79}$Rb (EC)→$^{79}$Kr | 0.35 | $\geq 2.3\times10^{20}$ y<br>22.9min | 129,147,183,688 | |
| p+$^{84}$Sr→$^{85}$Y (EC)→$^{85}$Sr | 0.5 | 2.68hrs,<br>$4.86^{(m)}$hrs | 231, 767 | 22 |
| p+$^{92}$Mo→$^{93}$Tc (EC, β)→$^{93}$Mo | 14.84 | 2.75hrs,<br>$43.5^{(m)}$min | 1363,1477,1520,<br>944, 1492 | 23,24 |
| p+$^{94}$Mo→$^{95}$Tc (EC, β)→$^{95}$Mo | 9.25 | 20hrs,<br>61dys | 766<br>204,786 | 23 |
| p+$^{96}$Ru→$^{97}$Rh (EC, β)→$^{97}$Ru | 5.54 | 30.7min,<br>46.2 min | 189,422,840,879,<br>528,771,2246 | 25 |
| p+$^{98}$Ru→$^{99}$Rh (EC, β)→$^{99}$Ru | 1.87 | 16.1dys,<br>4.7hrs | 89,322,618<br>340,1261 | 25 |
| p+$^{102}$Pd→$^{103}$Ag (EC, β)→$^{103}$Pd<br>→$^{103}$Ag (IT)→$^{103}$Ag | 1.02 | 65.7min,<br>5.7s | 118,244,267,532,<br>1274,134 | 26,27 |
| p+$^{106}$Cd→$^{107}$In (EC, β)→$^{107}$Cd<br>→$^{107}$In (IT)→$^{107}$In | 1.25 | 32.4min,<br>50.4s | 204,321,506,1268<br>678 | 28,29 |
| p+$^{108}$Cd→$^{109}$In(EC, β)→$^{109}$Cd<br>→$^{109}$In (IT)→$^{109}$In | 0.89 | $\geq 2.6\times10^{17}$ y<br>4.159hr,<br>1.34 min, | 203,<br>650 | 28,29 |
| p+$^{113}$In→$^{114}$Sn | 4.29 | Stable | No data | |
| p+$^{112}$Sn→$^{113}$Sb(EC, β)→$^{113}$Sn | 0.97 | 6.67min | 77(332),498 | 30, 31 |
| p+$^{114}$Sn→$^{115}$Sb(EC, | 0.66 | 32.1min,<br>6.2ns, 159ns | 497,<br>1300 | 32 |
| $\beta$)→$^{115}$Sn<br>→$^{115}$Sb (IT)→$^{115}$Sb | | | | |
| p+$^{115}$Sn→$^{116}$Sb(EC,<br>$\beta$)→$^{116}$Sn | 0.34 | 15.8 min,<br>60.3min | 1293,2225 | |
| p+$^{120}$Te→$^{121}$I (EC,<br>$\beta$)→$^{121}$Te | 0.09 | >2.2×10$^{16}$ y<br>2.12hrs | 212 | 33 |
| p+$^{124}$Xe→$^{125}$Cs(EC)→$^{125}$Xe<br>→$^{125}$Cs (IT)→$^{125}$Cs | 0.095 | 46.7min<br>≥1.1×10$^{17}$ y 0.90 ms | 112, 526,540,712,<br>77 (176),84(168) | |
| p+$^{126}$Xe→$^{127}$Cs(EC,<br>$\beta$)→$^{127}$Xe<br>→$^{127}$Cs (IT)→$^{127}$Cs | 0.089 | 6.25hrs<br>55 μs | 125, 412,<br>66 (386) | |
| p+$^{130}$Ba→$^{131}$La(EC,<br>$\beta$)→$^{131}$Ba<br>→$^{131}$La (IT)→$^{131}$La | 0.106 | 59min<br>≥3.5×10$^{14}$ y 170 μs | 108,285, 365,526<br>26(169) | |
| p+$^{132}$Ba→$^{133}$La(EC,<br>$\beta$)→$^{133}$Ba | 0.101 | >3.0×10$^{21}$ y<br>3.912hrs | 12(279),291,302,<br>858 | |
| p+$^{138}$La→$^{139}$Ce(EC,<br>$\beta$)→$^{139}$La<br>→$^{139}$Ce (IT)→$^{139}$Ce | 0.185 | 137.641dys,<br>1.02×10$^{11}$ y 57.58 s | 166*,<br>754 | |
| p+$^{136}$Ce→$^{137}$Pr(EC,<br>$\beta$)→$^{137}$Ce | 0.090 | >0.7×10$^{14}$ y 1.28hr | 160, 434, 514,763,<br>837, 867 | |
| p+$^{138}$Ce→$^{139}$Pr(EC,<br>$\beta$)→$^{139}$Ce | 0.251 | ≥0.9×10$^{14}$ y 4.41hr | 255,1320,1347, 1376,1631 | |
| p+$^{144}$Sm→$^{145}$Eu(EC,<br>$\beta$)→$^{145}$Sm | 3.07 | 5.93dys | 894,1658,1997 | |
| p+$^{152}$Gd→$^{153}$Tb(EC,<br>$\beta$)→$^{153}$Gd<br>→$^{153}$Tb(IT)→$^{153}$Tb | 0.20 | 1.08×10$^{14}$ y<br>2.34dys.<br>186 μs | 42(142),110,212,<br>81 | |
| p+$^{156}$Dy→$^{157}$Ho(EC,<br>$\beta$)→$^{157}$Dy | 0.06 | 12.6min | 61(87,280),188,341,<br>508,897,1211 | |
| p+$^{158}$Dy→$^{159}$Ho(EC,<br>$\beta$)→$^{159}$Dy<br>→$^{159}$Ho(IT)→$^{159}$Ho | 0.10 | 33.05min,<br>8.30 s | 57(121,253),178,310,166,2<br>06 | |
| p+$^{162}$Er→$^{163}$Tm(EC,<br>$\beta$)→$^{163}$Er | 0.139 | 1.81hrs | 69(371,393),104 | |
| p+$^{164}$Er→$^{165}$Tm(EC,<br>$\beta$)→$^{165}$Er | 1.601 | 30.06hrs | 47(196),243,296,297 | |
| p+$^{168}$Yb→$^{169}$Lu(EC,<br>$\beta$)→$^{169}$Yb<br>→$^{169}$Lu(IT)→$^{169}$Lu | 0.13 | 34.06hrs,<br>160s | 191,961,1450,<br>29 | |
| p+$^{174}$Hf→$^{175}$Ta(EC,<br>$\beta$)→$^{175}$Hf | 0.16 | 2.0×10$^{15}$ y 10.5hrs | 82(104,126,267),126,207,3<br>49,475,1226,<br>1793 | |
| p+$^{180}$Ta→$^{181}$W(EC)→$^{181}$Ta | | 8.154 h<br>>1.2×10$^{15}$ y<br>121.2dys | 6(152) | |
| p+$^{180}$W→$^{181}$Re(EC, | 0.12 | 1.8×10$^{18}$ y | 366 | |
| β)→$^{181}$W | | 19.9hrs | | |
| p+$^{184}$Os→$^{185}$Ir(EC)→$^{185}$Os | 0.02 | >5.6×10$^{13}$ y<br>14.4hrs | | 37(161),97(254),222 |
| p+$^{190}$Pt→$^{191}$Au(EC, β)→$^{191}$Pt →$^{191}$Au(IT)→$^{191}$Au | 0.014 | 6.5×10$^{11}$ y<br>3.18hr. | 0.92s | 167,254,278,293, 400,488,253 |
| p+$^{196}$Hg→$^{197}$Tl(EC, β)→$^{197}$Hg →$^{197}$Tl(IT)→$^{197}$Tl | | | 2.84hrs<br>0.54s | 134,152,308,309,578, 793,386 |

**TABLE 2:** Some experimental parameters from References in Table 1:

| p-nuclei | Proton energy (MeV) | Target& Backing | Method | γ-Ray Detector | Beam Current |
|---|---|---|---|---|---|
| | | | | | |
| $^{74}$Se | 1.3 -3.6 | Nat. Se (thick Al) 200-700 $\mu g/cm^2$ | Activation | Single HPGe | 5-10 $\mu$A |
| $^{78}$Kr | | | | | |
| $^{84}$Sr | 1.5 – 3 | Nat. SrF2 (thick C) | Activation | Single HPGe | 5-10 $\mu$A |
| $^{92}$Mo | 1.5 – 3 | Nat. Mo (1mm Al) 0.12 – 0.5 $\mu m$ | Activation | Single HPGe | 45 $\mu$A |
| | Gamow region | Enriched $^{92}$Mo 400 $\mu g/cm^2$ | In-beam | 13 HPGe(HORUS) | 200$n$A |
| $^{94}$Mo | 1.5 - 3 | Nat. Mo (1mm Al) 0.12 – 0.5 $\mu m$ | Activation | Single HPGe | 45 $\mu$A |
| $^{96}$Ru | 1.5-3.5 | Nat. Ru(1mm Al) | Activation | Single HPGe | 20-50 $\mu$A |
| $^{98}$Ru | 1.5-3.5 | Nat. Ru(1mm Al) | Activation | Single HPGe | 20-50 $\mu$A |
| $^{102}$Pd | 2.68–6.85 | Nat. Pd(Al 1mm Al) 420-520$nm$ | Activation | Single HPGe | 10 $\mu$A |
| $^{106}$Cd | 2.4-4.8 | Nat Cd and enriched (96.47%) (3$\mu m$Al) 100-600 $\mu g/cm^2$ | Activation | Single HPGe | 500$n$A |
| $^{108}$Cd | 2.4-4.8 | Nat Cd and enriched (2.05%) (3$\mu m$Al) 100-600 $\mu g/cm^2$ | Activation | Single HPGe | 500$n$A |
| $^{112}$Sn | 3-8.5 | Enriched (98.9%) 2.7 $mg/cm2$ | Activation | Two BGO (one HPGe for energy calibration) | 100-150 $n$A |
| $^{114}$Sn | 1-3.7 | Enriched (71.1%) 0.05 $mg/cm2$ | Activation | Two coaxial HPGe | |
| $^{120}$Te | 2.5-8 | Enriched (99.4%) 128$\mu g/cm^2$(20$\mu g/cm^2 C$) 456$\mu g/cm^2$(1.5$mg/cm^2 C$) | Activation | Two Clover detectors | 80-320$n$A |

**B. Discussion (Tables 1, 2; Figure 1)**

In the present study, we focused on previous studies where stable *p*-nuclei were bombarded with a proton beam and the capture cross-sections were measured by detecting the decaying gamma rays. As seen from Table 1, the abundances of these *p*-nuclei are considerably low with the exception of some of them, such as Mo, Ru, In, and Sm. In a proton capture reaction experiment, in which the *p*-nuclei are being used as the target, low abundance results in very low yield, necessitating high beam current and prolonged activation to obtain conceivable statistics for deducing the reaction cross-sections with considerable certainity. The half lives of the daughter nuclei are small (of the order of minutes or 1−2 hrs.) in most cases, making the detection process very difficult or even impossible. Systems with comparatively large half lives for the daughter nuclei, *viz.* $^{98}$Ru, $^{108}$Cd, $^{126}$Xe, $^{132}$Ba, $^{138}$Ce, $^{144}$Sm, $^{152}$Gd, $^{164}$Er, $^{168}$Yb, $^{174}$Hf and $^{180}$Ta, can be more easily studied using the activation technique. The nuclei, which exhibit much larger half lives (of the order of many days or years), are generally studied using the in-beam technique. The fourth column of Table 1 lists the gamma rays to be detected for measuring the reaction cross-sections. The intensities of very low energy gamma rays, which cannot be detected owing to experimental limitations, can be deduced by measuring intensities of the higher energy transitions (in parenthesis) leading to the low energy lines after correcting for the corresponding branching transitions. From Table 2, we observe that the majority of the systems were experimentally studied using the activation method and a single HPGe detector. Both natural and enriched targets were used either as thick, self-supporting or thin, backed. The most common backing material used is Al. The use of Al as backing material, having much lower Z than the target material, helps avoid interferences from the events occurring from the interaction between the projectile and backing. The beam current used in all the experiments was reasonably high to achieve sufficient yield. The system $p$+$^{92}$Mo was studied in both activation and in-beam methods. For in-beam studies [12], an enriched self-supported target and 13 HPGe detectors were used, whereas for activation measurements, [23] backed target and a single HPGe detector were used. The cross section results from in-beam studies [12] were compared with earlier activation measurements [23] and they were found to be in reasonable agreement, except for a few discrepancies. Figures 1 (a) and (b) represent the variation of the Coulomb barrier and *Q*-value of the reactions against mass numbers of the *p*-nuclei. The variation of Gamow peaks with corresponding mass numbers are represented in Fig. 1(c), in which the Gamow window is denoted as error bars. Note from Fig. 1 (a) that, as expected, the Coulomb barrier increases with the mass number. The *Q* value (Fig. 1 (b)) was found to be within the range of 3−5 MeV with the exception of $^{113}$In, $^{138}$La, and $^{180}$Ta, for which the value is within ~7−8 MeV. The Gamow window was found to be within the range of ~0.5 to 7.5 MeV.

**C. TALYS calculations**

TALYS is a computer code system for analysis and prediction of nuclear reactions [33]. It is a nuclear physics tool that can be used for the analysis of nuclear reaction experiments, either in a default mode when no measurements are available, or after finely tuning the adjustable parameters of the various reaction models using available experimental data. Rather than providing a detailed description of only one or a few reaction channels, TALYS outputs a simultaneous prediction of many nuclear reaction channels. TALYS is meant for the analysis of data in the energy range of 1 keV−200 MeV. The lower limit is the energy above the resolved resonance range, which is not necessarily exactly 1 keV. In this study, TALYS calculations were performed to obtain theoretical estimates of the proton-capture cross section values. The results are represented against the proton energy in Figs. 2 and 3. The corresponding experimentalvalues obtained from the KaDONiS [16] database are included for comparison in Fig. 2. The corresponding values given in the TENDL tabulation [34] are also included. TENDL is the commonly referred nuclear data library that provides the output of the TALYS nuclear model code system. It was reported by the authors [33] that for protons, libraries produced by default TALYS calculations perform reasonably well. The present TALYS calculations were performed with the default parameters [33]. TALYS calculations were also performed for all the other *p*-nuclei for which no experimental data have been reported. These results are represented in Fig. 3. The plots also include the corresponding results from TENDL tabulation. As mentioned above, TALYS can generate nuclear data for all open reaction channels after finely tuning the adjustable parameters of the various reaction models using available experimental data. TALYS incorporates a completely integrated optical model and recent optical model parameterizations for many nuclei. The central assumption of this optical model is that the complicated interaction between an incident particle and a nucleus can be represented by a complex mean-field potential, which divides the reaction flux into two parts: one representing the shape elastic scattering and the other representing all competing non-elastic channels. The phenomenological OMP for nucleon-nucleus scattering, *U*, is defined as [33]

$$U(r,E) = -V_V(r,E) - iW_V(r,E) - iW_D(r,E) + V_{SO}(r,E).l.\sigma + iW_{SO}(r,E).l.\sigma + V_C(r) \quad (5)$$

where $V_{V,SO}$ and $W_{V,D,SO}$ are the real and imaginary components of the volume-central ($V$), surface central ($D$), and spin-orbit ($SO$) potentials, respectively. $E$ is the energy of the incident particle in MeV in the laboratory frame. All components are separated in energy-dependent parts ($VV$, $WV$, $WD$, $VSO$, and $WSO$) and energy–independent radial parts $f$ [33]. The form factor $f(r, R_i, a_i)$ is a Woods-Saxon shape expressed as

$$f(r,R_i,a_i) = (1 + \exp[(r - R_i)/a_i])^{-1} \tag{6}$$

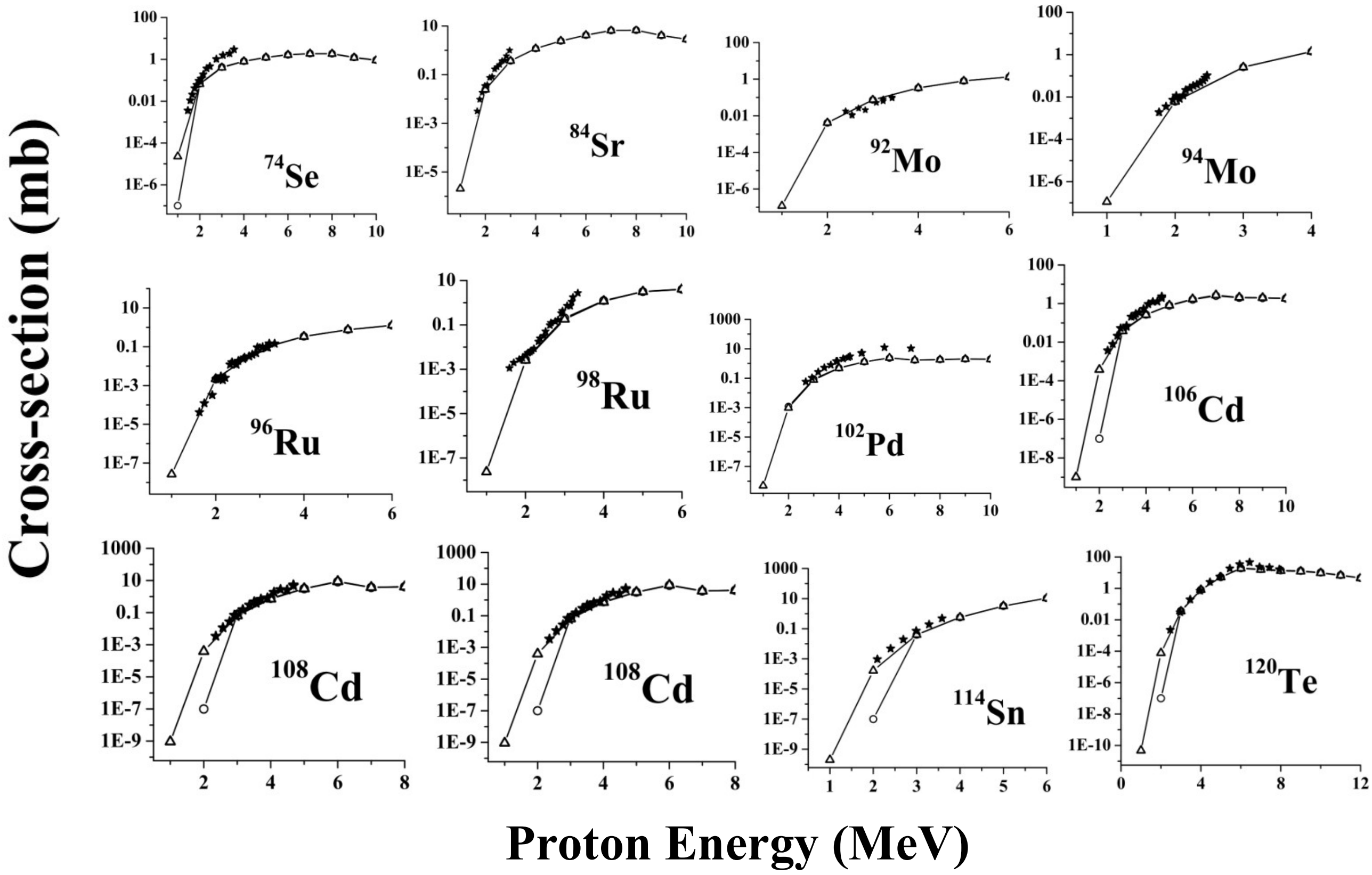

**Fig. 2.** Plots of the proton capture cross-section values vs. energy in the Gamow region. Experimental values (solid star), TALYS calculation results (open circle), and results from TENDL tabulation (open triangle) are compared. The theoretical values are connected by straight lines to facilitate visual analysis.

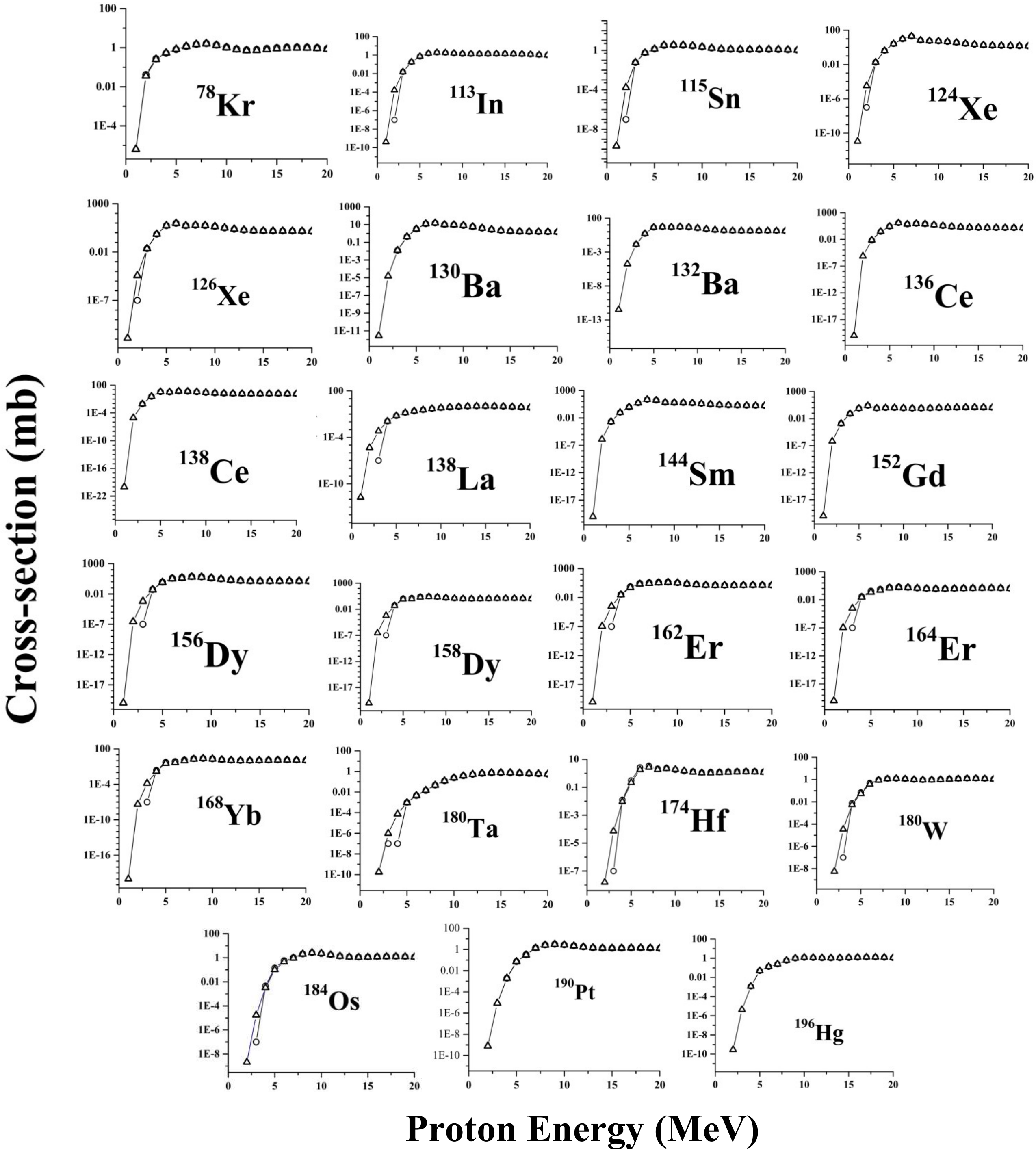

**Fig. 3.** Plots of the cross-section values from TALYS calculation (open circle) and TENDL tabulation (open triangle) for the systems for which no experimental data have been reported. The theoretical values are connected by straight lines to facilitate visual analysis

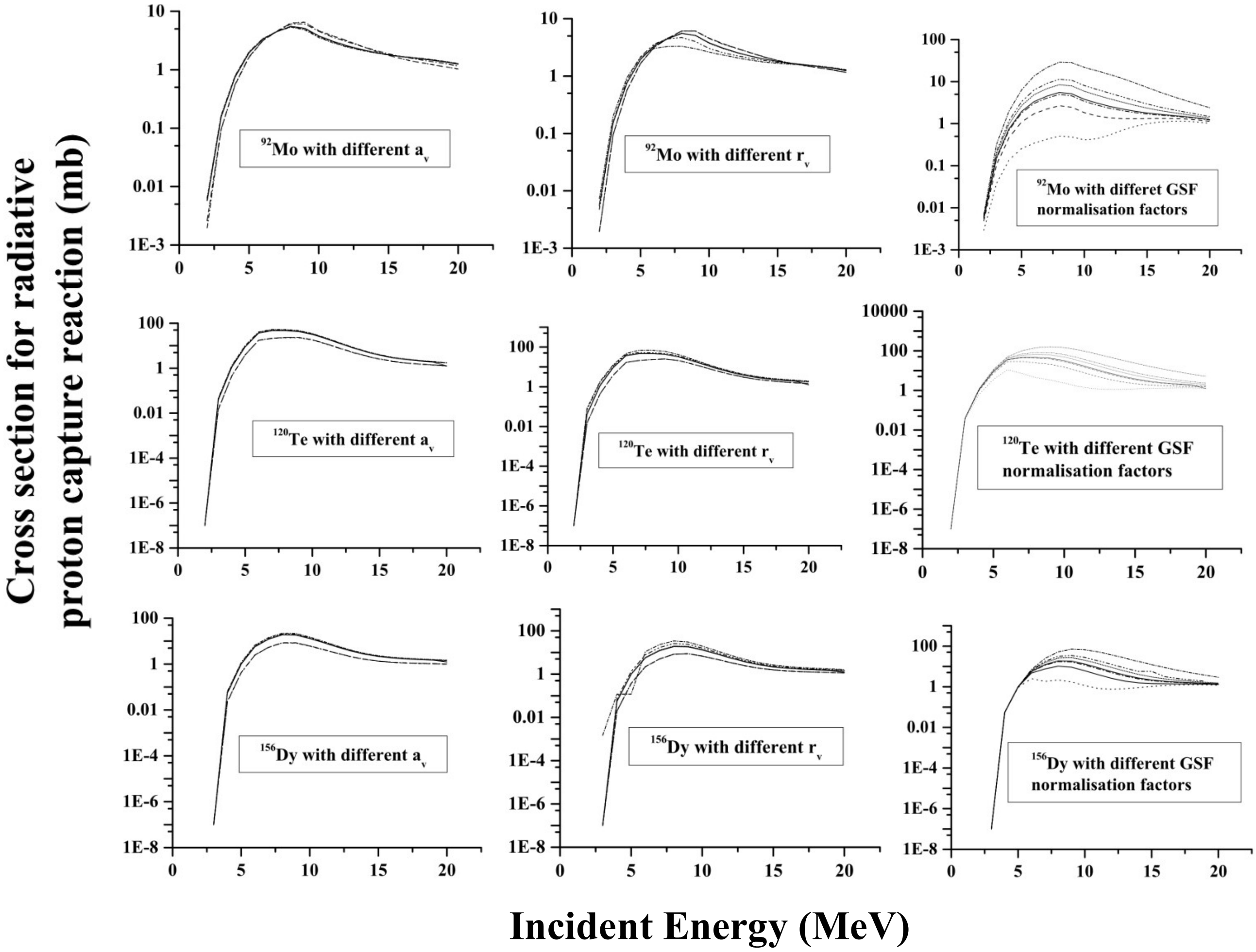

**Fig. 4:** Plots of the cross-section values from TALYS calculations with different values of $a_v$, $r_v$, and *GSF* for $^{92}$Mo, $^{120}$Te, and $^{156}$Dy.

The default optical model potentials (OMPs) used in TALYS are the local and global parameterizations of Koning and Delaroche [35]. Gamma-ray transmission coefficients are important for radiative capture reactions involving gamma emission. These coefficients depend on the gamma–ray energies and gamma-ray strength functions (GSF) [33]. In the present study, we performed a few calculations by changing the OMPs, $a_v$ and $r_v$ ,as well as the GSF for three nuclei, namely $^{92}$Mo, $^{120}$Te, and $^{156}$Dy, to obtain a preliminary idea of how they affect the cross-section values. The $r_v$ value was varied within the range from approximately 0.12 to 0.15; and the corresponding default value is approximately 1.2. The $a_v$ value was changed within the range from approximately 0.06 to 0.73; the corresponding default value is approximately 0.67. The Gamma-strength normalization factors were varied from 0.1 to 1.1. The corresponding plots are shown in Fig. 4. Note from this figure that, within the Gamow energy region(2−5 MeV), the cross-section values apparently do not change significantly with GSF, $a_v$, and $r_v$. The variation becomes more pronounced as we move to higher energies. These data are not sufficient to make any conclusive comments about the dependence of the cross-section values on OMPs and GSFs. This is work in progress towards a more elaborate and detailed study for a potential separate publication.

## IV. SUMMARY

The present study aims at providing a guide to study the p-nuclei by presenting a short review of the present status of the investigations in this field and estimates of radiative proton capture cross-sections obtained from TALYS calculations. It was observed that in the majority of the experimental investigations, a single HPGe detector was used to detect the gamma rays. As mentioned earlier, and according to Fig. 2, the proton capture cross-sections are generally very small. Hence, more powerful detectors such as Clover can be used to improve the detection efficiency. Table 2 shows that natural targets were used in most experiments. The presence of isotopes other than the one of primary interest gives rise to events that often eclipse the event of interest. To improve the relative yield of the gamma rays of interest, experiments may be repeated with enriched targets in systems where previous investigations were performed using natural targets. It is evident from the present review that there are very few *p* nuclei above a mass $A$~100, for which experimental cross-section data have been reported. There is a notable demand of experimental investigations on *p*-nuclei. Most of the systems were investigated through the activation technique. For these systems, where reliable data exist from activation studies, in-beam measurements may be performed. This will help study those systems for which activation studies are not viable and identify and correct for the beam induced background and angular distribution effects, which are major sources of errors in in-beam investigations.

The TALYS calculations were performed to obtain theoretical estimates of the proton capture cross-section values on the *p*-nuclei. These values were compared with the corresponding experimental data reported in the literature and data from the TENDL tabulation. Although the experimental data seem to agree well with the theoretical estimates within the Gamow region, we should keep in mind that the cross-section values in this region are extremely small, approximately few μb or even smaller. Hence, even very little differences between experimental and theoretical values cannot be ignored. Note that both theoretical and experimental cross-section values show a similar nature of rise and fall with the change in the incident energy. As we move to higher energy, the differences between experimental and theoretical values appear to become more pronounced. All the aforementioned calculations were performed using default parameter values. Preliminary results are presented, depicting the behaviour of the cross-section values over a short range of few optical model parameters and gamma strength functions. This is work in progress; calculations over a wider range for all the parameters and all *p*-nuclei are expected to provide sufficient predictive power to the various nuclear models and give an indication of the reliability of the measurements

## Acknowledgement

One of the authors, Indrani Ray, is indebted to Prof. Mitali Mondal (Physics Department of Jadavpur University) for her help and support. The same author is immensely grateful to Prof. Maitreyee Saha Sarkar (retired from Saha Institute of Nuclear Physics, Kolkata) for her continuous guidance, and thankful to Prof. Polash Banerjee (Retired Professor, SINP) for his guidance during the initial part of project planning. The same author is also indebted to Prof. Chinmoy Basu (Nuclear Physics Group), Prof. Sukalyan Chattopadhyay (Head, Nuclear Physics Group), and the Director of Saha Institute of Nuclear Physics (SINP) for their support. The authors are also thankful to Dr. Sathi Sharma (Ex-Ph D student of SINP) and Dipali Basak (Ph D student, SINP) for their help at various stages of this work.